# Logic Programming, Functional Programming, and Inductive Definitions

*Lawrence C. Paulson* 1 *and Andrew W. Smith* 2

1 Computer Laboratory, University of Cambridge, Cambridge CB2 3QG, England
2 Harlequin Limited, Barrington, Cambridge CB2 5RG, England

**Table of Contents**

# 1 Introduction

The unification of logic and functional programming, like the Holy Grail, is sought by countless people [6, 14]. In reporting our attempt, we first discuss the motivation. We argue that logic programming is still immature, compared with functional programming, because few logic programs are both useful and pure. Functions can help to purify logic programming, for they can eliminate certain uses of the cut and can express certain negations positively.

More generally, we suggest that the traditional paradigm — logic programming as first-order logic — is seriously out of step with practice. We offer an alternative paradigm. We view the logic program as an *inductive definition* of sets and relations. This view explains certain uses of Negation as Failure, and explains why most attempts to extend Prolog to larger fragments of first-order logic have not been successful. It suggests a logic language with functions, incorporating equational unification.

We have implemented a prototype of this language. It is very slow, but complete, and appear to be faster than some earlier implementations of narrowing. Our experiments illustrate the programming style and shed light on the further development of such languages.

# 2 Declarative programmers: realists versus purists

Logic Programming and Functional Programming are often lumped together under the heading 'Declarative Programming'. Ideally, a declarative program simply specifies the problem — what we want — and the computer works out how to do it.

Of course this is an oversimplification. For the declarative languages that exist now, the problem description really is a program: not for any physical machine, but for an abstract machine. A functional program defines a system of rewriting rules that can evaluate a desired function. A logic program defines a search space of problem reductions that can solve all instances of the desired goal. The declarative program expresses the algorithm more abstractly than, say, a Pascal program, but the means of expression are restrictive when regarded as a specification language: even more so if we care about efficiency.

Users of declarative languages can be described as *realists* or *purists*:

- Realists set out to write useful programs. While they value the declarative reading, they are prepared to compromise it if necessary.
- Purists set out to demonstrate their declarative paradigm, and perhaps its application to program correctness and synthesis. Their programs are completely pure, regardless of the consequences for efficiency.

Realists and purists are equally worthy; they simply have different priorities. Note that an impure program can be more readable than a pure one, while a pure program can be more efficient than an impure one.

Let us compare functional programming and logic programming through these concepts. For the purist view we can compare the presentations by David Turner and Robert Kowalski at a special meeting of the Royal Society in London.

### 2.1 Functional programming

A purist functional programmer might use Miranda, Lazy ML, or Haskell: lazy functional languages with no side-effects whatever. David Turner's presentation to the Royal Society includes quick sort, a topological sort, and a program to find a Knight's tour of the chess board [42]. He also gives some simple proofs and program derivations. Bird and Wadler [7] give a fuller account of the purist approach; they derive functional programs from formal specifications.

Purists avoid LISP because of its imperative features. David Turner says [42, page 53]:

> It needs to be said very firmly that LISP, at least as represented by the dialects in common use, is not a functional language at all.

LISP has been impure from the very start. Assignments and go to's feature prominently in the LISP *1.5 Programmer's Manual* [32]. But that same book devotes a chapter to a program for the propositional calculus (Wang's Algorithm). This is a substantial, purely functional program — probably the first ever — and it is written in LISP.

A realist functional programmer might use LISP or ML. These languages support a functional style but do not enforce it. Abelson and Sussman [1] illustrate the realist approach using Scheme, a dialect of LISP. Many of their examples are purely functional.

The realists and the purists share some common ground. Many LISP and ML programmers strive for a pure style, while many pure functional programs can be executed with reasonable efficiency.

### 2.2 Logic programming

Kowalski illustrates the purist approach. His presentation to the Royal Society emphasises the relationship between logic programs and specifications [28, page 11]:

> The only difference between a complete specification and a program is one of efficiency.

As an example, Kowalski gives a specification of sorting. He then gives the following sorting program:

```
sort(X,X) :- ordered(X).
sort(X,Y) :- I<J, X[I] > X[J], interchange(X,I,J,Z), sort(Z,Y).
```

Given `X`, the program computes the sorted version `Y` by repeatedly exchanging some `X[I]` and `X[J]` that are out of order. The program is highly nondeterministic: the condition `X[I] > X[J]` is the only constraint on `I` and `J`.[3] To regard this as a useful sorting program we must further constrain `I` and `J`, to reduce the search drastically. We also must find a compiler clever enough to execute `interchange(X,I,J,Z)` without copying.

Perhaps it is unfair to bear down on this little example. But the literature offers few others. Hogger [26] writes at length about pure logic programs, typically to reverse a list or test for list membership. By comparison, the pure functional programs in Bird and Wadler [7] perform $\alpha$-$\beta$ search, construct Huffman coding trees, and print calendars.

Clocksin and Mellish [13] illustrate the realist approach. They teach PROLOG style using a wide variety of programs, with applications such as parsing. But many of these involve logically meaningless (or 'extralogical') predicates.

For logic programming, the realists and purists are far apart. Programming in a pure style is difficult. Existing PROLOG systems do not even provide pure PROLOG as a subset. They use depth-first search (which is incomplete) and they omit the occurs check (which can create circular data structures).

Pure logic programs can be written by translating functional programs into clauses. But this is hardly logic programming: key aspects like backtracking are lost. Logic programming is far more ambitious than functional programming, which is why it has not reached a similar stage of maturity.

The widespread interest in extending PROLOG stems mainly from purist principles. Kowalski again [28, page 22]:

> In the longer term, we need to develop improved logic programming languages, which do not rely on extralogical features for the sake of efficiency.

## 3 Logic programs: first-order theories or inductive definitions?

We need an improved logic programming paradigm, not just an improved language, if pure logic programming is to become practical. So let us consider what logic programming really means. We begin with the orthodox view and then propose an alternative.

---

3 The subscripting in `X[I] > X[J]` abbreviates `contains(X,I,U), contains(X,J,V), U>V`

### 3.1 Logic programs as first-order theories

PROLOG is descended from Robinson's resolution principle for proving theorems in first-order logic [37]. Every clause in a pure PROLOG program stands for a first-order formula; we are PROgramming in LOGic. To illustrate this orthodox view, consider the traditional 'family relationships' example:

```
grandparent(X,Z) :- parent(X,Y), parent(Y,Z).
cousin(X,Y) :- grandparent(Z,X), grandparent(Z,Y).
parent(elizabeth,charles).
parent(elizabeth,andrew).
parent(charles,william).
parent(charles,henry).
parent(andrew,beatrice).
```

We can regard this PROLOG program as a first-order theory. The first clause corresponds to the logical axiom

$$(\forall x)(\forall y)(\forall z)\,\mathrm{parent}(x,y) \wedge \mathrm{parent}(y,z) \rightarrow \mathrm{grandparent}(x,z)$$

If we pose the query

```
?- cousin(henry,beatrice).
```

then PROLOG answers `yes`, for the query has an obvious proof from the axioms.

If we ask

```
?- cousin(elizabeth,asterix).
```

then PROLOG answers `no`, for the query has no proof. But it has not been disproved, for it is true in some models, false in others. The answer `no` conveys less information than `yes`.

Now imagine we pose the negative query

```
?- not cousin(elizabeth,asterix).
```

A typical PROLOG system will answer `yes` because there is no proof of

```
cousin(elizabeth,asterix).
```

This treatment of negation is called *Negation as Failure*; it differs from logical negation, since the query is not a logical consequence of the axioms. Some people think that Negation as Failure is a cheap hack, and that researchers should aim to implement logical negation. Logical negation would answer `no` to our query — but this is often undesirable. PROLOG programmers recognise that our database *defines relations* such as 'cousin-of', where `cousin(elizabeth,asterix)` does not hold. Negation as Failure is a natural way to test whether a relation holds. This point of view is hard to justify under the orthodox paradigm for logic programming.

### 3.2 Logic programs as inductive definitions

Here is an alternative paradigm for logic programming. A set of clauses is not a first-order theory, but the definition of a new logic. The meaning of a logic program is the set of theorems in this 'private' logic: all derivable ground atoms. This is a monotone inductive definition of a family of sets, one for each predicate.

Inductive definitions appear in all branches of mathematics. The natural numbers are the least set containing 0 and closed under successor. The Boolean expressions are the least set containing propositional letters and closed under $\wedge$, $\vee$, $\neg$. Most importantly: the set of *theorems* in a logical system is the least set containing all axioms and closed under all applications of inference rules. As Aczel explains, this is the general form of an inductive definition [2].

What has this to do with logic programming? We can regard a logic program as an inductive definition by taking its clauses as axioms and inference rules. We regard our family relationships database as a new logic with rules like

$$\frac{\text{parent}(x,y) \quad \text{parent}(y,z)}{\text{grandparent}(x,z)} \qquad \frac{\text{grandparent}(z,x) \quad \text{grandparent}(z,y)}{\text{cousin}(x,y)}$$

This inductively defines various sets. The 'grandparent of' relation is the set of all pairs $\langle x,y \rangle$ such that grandparent$(x,y)$ follows from the database. Similarly the derivable instances of cousin$(x,y)$ define the 'cousin of' relation.

Aczel [2] gives the semantics of an inductive definition as follows.

- A *rule* has the form $p \leftarrow P$, where $P$ is the set of premises and $p$ is the conclusion.
- Let $\Phi$ be a set of rules. A set $A$ is $\Phi$-*closed* provided that for each rule $p \leftarrow P$ in $\Phi$, if $P \subseteq A$ then $p \in A$. (Thus if the premises are in $A$ then so is the conclusion.)
- The set $I(\Phi)$ *inductively defined* by $\Phi$ is given by

$$I(\Phi) = \bigcap \{A \mid A \text{ is } \Phi\text{-closed}\}$$

The inductively defined set $I(\Phi)$ can also be expressed as the least fixed point of a monotone operator. A set of rules $\Phi$ defines a universe or assertion language $A$, namely the set of all premises and conclusions of rules:

$$A = \bigcup \{P \cup \{p\} \mid \text{rule } p \leftarrow P \text{ is in } \Phi\}$$

Note that $A$ corresponds to the Herbrand base of a set of clauses. Now $\Phi$ defines a monotone operator $\phi$ over $A$, corresponding to all possible rule applications in a set $Y$ of assertions. Precisely, if $Y \subseteq A$ then

$$\phi(Y) = \{p \in A \mid \text{rule } p \leftarrow P \text{ is in } \Phi \text{ and } P \subseteq Y\}$$

The iterates of $\phi$ are defined as usual:

$$\phi^0 = \emptyset$$
$$\phi^{n+1} = \phi^n \cup \phi(\phi^n)$$
$$\phi^\omega = \bigcup_{n \in \omega} \phi^n$$

If all rules in $\Phi$ have a finite number of premises then $\phi^\omega = I(\Phi)$ and $\phi^\omega$ is the least fixed point of $\phi$.

The full theory of inductive definitions is complicated, but much of it need not concern us. A rule $p \leftarrow P$ could have an infinite number of premises, unlike rules in logic programs. The rules in an inductive definition contain no variables. A schematic rule (like the rule for `cousin`) abbreviates an infinite set of rules: all ground instances under the Herbrand universe.

In the semantics of logic programming, such theory has long been used as a technical device. We suggest, rather, that an inductive definition is a logic program's intrinsic declarative content. Clauses should not be viewed as assertions in first-order logic, but as rules generating a set.

For a concrete example, consider how a formal grammar generates the strings of a language. Grammars are inductive definitions. This may explain why logic programming works so well at natural language processing.

### 3.3 Least models and the Closed World Assumption

Definite clauses are a fragment of first-order logic enjoying remarkable qualities. Any set of definite clauses is consistent. In the greatest model, each predicate is universally true; in the unique least model, each predicate holds just when it must. The least model is the interesting one, for it corresponds to our intuition that our logic program defines a set of relationships.

Van Emden and Kowalski [43] observed that the model-theoretic semantics of a logic program is best given by the least Herbrand model, which is the intersection of all Herbrand models. This coincides with the operational and fixed point semantics. Their fixed point semantics is precisely our inductively defined set $I(\Phi)$.

The least model can be formalized in first-order logic as the Closed World Assumption, augmenting the database with the negations of all ground atoms that do not hold in the least model. Shepherdson shows how this leads to difficulties [40]. First-order logic is simply too weak to characterize the least model.[4] Horn clause logic is even weaker. But the least model can be directly expressed by an inductive definition.

---

[4] By the Skolem-Löwenheim Theorem, no set of first-order axioms can even fix the cardinality of its models, let alone fix a single model.

Negation as Failure is investigated by Apt and van Emden [3] and Lloyd [29]. Essentially, they develop the theory of inductive definitions so as to distinguish divergent computations from finite failures. Negation goes beyond monotone inductive definitions: with negated subgoals, the function $\phi$ above may not be monotone. However, perhaps the database can be partitioned into several inductive definitions, so that each negation refers to a set that has already been defined (the dependency graph must be acyclic). The database can then be interpreted as an iterated inductive definition (via some treatment of finite failure.) Such databases are called *stratified* or *free from recursive negation* [44]. The main stream of (sound) research into negation [34] uses the mathematics of fixedpoints, ordinals, and inductive definitions, not that of classical first-order logic.

In different situations, either view of logic programming — inductive definitions or first-order logic — could be more useful. Where the Closed World Assumption is wrong, so is the inductive view. Below we contrast these views with respect to several aspects of logic programming.

### 3.4 Specification and verification of logic programs

A key selling point for 'programming in logic' is that programs can be viewed as specifications. Programs can be derived from specifications, and these programs are guaranteed to be correct. Can we verify logic programs when regarding them as inductive definitions?

Given a specification, a logic program is *correct* if it is sound and complete. *Sound* (or partially correct) means that each successful goal in an execution is permitted by the specification. *Complete* means that each permitted goal will succeed during execution. Specifications are still written in some sort of logical formalism even when we regard programs as inductive definitions.

Every inductive definition gives rise to a principle of inductive proof. In simple cases, this principle resembles structural induction or mathematical induction (on the natural numbers). The general principle is induction over derivations in our 'private' logic. It is used to prove soundness of logic programs, basically by showing that each rule is individually sound. Completeness proofs typically involve some form of induction over the data, showing that the rules suffice for all the necessary derivations.

Fitting gives examples of correctness proofs [17, pages 49–53]. His book is a unique treatment of computability theory in the context of logic programming. He presents logic programs not as first-order theories, but as 'elementary formal systems', which are a restricted case of inductive definitions.

Kowalski [28, page 19] says that a program is totally correct provided it is logically equivalent to its specification. That is nice and simple. But Hogger notes [26, page 141]

> In practice, though, this is rarely possible: logic procedure sets usually have less information content than the specifications to which they conform, even though they may be complete.

Hogger goes on to demonstrate a more general method called *definiens transformation*. Each predicate is defined as logically equivalent to some formula. These definitions are transformed by replacing formulae by equivalent formulae. Finally each 'if-and-only-if' is replaced by 'if' when this results in a set of definite clauses. Hogger states [26, page 153]

> If this is accomplished successfully then the program (P,G) is thereby proven to be totally correct. Its partial correctness is directly established .... Its completeness is established by the fact that each definiens transformation preserves equivalence ...

Let us illustrate this method by deriving a predicate $\text{nat}(X)$ meaning '$X$ is a natural number'. Informally, let us say that $X$ is a natural number if and only if $X$ has the form $s(s(\cdots s(0) \cdots))$.

The specification is

$$\text{nat}(X) \leftrightarrow X \text{ is a natural number.}$$

Now $X$ is a natural number if and only if $X = 0$ or $X$ is the successor $s(Y)$ of some natural number $Y$. Therefore we may transform the specification to

$$\text{nat}(X) \leftrightarrow X = 0 \vee (\exists Y)(X = s(Y) \wedge Y \text{ is a natural number}).$$

Substituting back the original specification introduces a recursive call:

$$\text{nat}(X) \leftrightarrow X = 0 \vee (\exists Y)(X = s(Y) \wedge \text{nat}(Y))$$

Each disjunct has the form of a clause body, so replace $\leftrightarrow$ by $\leftarrow$:

$$\text{nat}(X) \leftarrow X = 0 \vee (\exists Y)(X = s(Y) \wedge \text{nat}(Y))$$

Simplification yields two clauses, a correct program:

$$\text{nat}(0) \qquad \text{nat}(s(Y)) \leftarrow \text{nat}(Y)$$

But here is another derivation. Note that $X$ is a natural number if and only if $s(X)$ is a natural numer. Therefore the specification is equivalent to

$$\text{nat}(X) \leftrightarrow s(X) \text{ is a natural number.}$$

Substituting back the original specification introduces a recursive call:

$$\text{nat}(X) \leftrightarrow \text{nat}(s(X))$$

Dropping the $\leftrightarrow$ gives the clause

$$\mathrm{nat}(X) \leftarrow \mathrm{nat}(s(X)).$$

How did we get such a useless program? Hogger's verification method requires a separate termination proof, which must be performed with respect to a given computational strategy [26, page 143–150]. So the connection between 'logic program' and 'logic specification' is not as simple as commonly thought.

When the above programs are viewed as inductive definitions, it is obvious that the first defines the natural numbers and the second defines the empty set. We can understand the 'definiens transformations' as equational reasoning on sets. If a set satisfies a recursive equation like $S = f(S)$, then $S$ is some fixedpoint of $f$. Replacing $\leftrightarrow$ by $\leftarrow$ picks out the least fixedpoint. These fixedpoints could differ, as they did above. The theory of inductive definitions could lead to better techniques [16].

## 4 Extended logic programming languages

Many extended logic programming languages aim to increase the power of pure declarative programming. Most extensions adopt the first-order logic viewpoint, but several are best understood from the viewpoint of inductive definition. While surveying other work, this section also discusses the design of our logic language with functions.

### 4.1 Larger fragments of first-order logic

If our goal is to program in logic then we should go beyond Horn clauses, aiming ultimately at programming in full first-order logic. Bowen [10] proposed a complete theorem-prover where programs consist of sequents of the form $A_1, \ldots, A_m \vdash B_1, \ldots, B_n$; a standard PROLOG interpreter handles the case where these resemble definite clauses. Many similar proposals have appeared since.

Stickel's Prolog Technology Theorem Prover [41] exploits the sophistication of current PROLOG implementations. He extends them to full first-order logic using sound unification with occurs check, the model-elimination inference rule, and depth-first iterative deepening for completeness. Stickel's stated aim is high-performance theorem proving; he specifically de-emphasises its potential for logic programming [41, page 375].

Applications of a full first-order logic programming language are hard to visualize. Perhaps the problem is that first-order logic destroys that vital property, the least model property. The disjunction $p \vee q$ has two minimal models, where either $p$ or $q$ is true. Therefore, disjunctive axioms destroy the least model property. So do negative goals and nested implication, for $p \leftarrow \neg q$ and $q \leftarrow (q \leftarrow p)$ are classically equivalent to $p \vee q$. Makowsky has formalized the least model property in terms of initial structures and generic examples. He shows that Horn clauses are the largest fragment of first-order logic enjoying this property [30]. By regarding logic programs as inductive definitions, we come to the same conclusion at once. Makowsky's work is rigorous confirmation of our intuitive idea.

### 4.2 Other work concerning inductive definitions

Hagiya and Sakurai [23] present a formal system for logic programming, based on the theory of iterative inductive definitions. This system captures the least fixedpoint semantics of a set of clauses and formally justifies Negation as Failure. They envisage programs consisting of several levels, each defined inductively in terms of its predecessor. The formal system is given as a foundation for PROLOG, with applications to program specification, verification, and synthesis. Hagiya and Sakurai take the traditional first-order view of logic programming, but at times appear to question this paradigm [23, page 71]:

> PROLOG usually explained as being based on SLD-resolution. It is more natural, however, to regard a PROLOG program and its execution as a set of productions and generation of a normal proof than to regard them as a set of Horn clauses and SLD-resolution, since it more faithfully reflects the procedural interpretation of predicate logic ... Some resolution procedures are more clearly understood in terms of deduction, even if deduction and refutation are equivalent.

Hallnäs and Schroeder-Heister [25] advance a view of logic programming based on inductive definitions. Calling the traditional view 'clauses-as-formulae', they advocate instead 'clauses-as-rules': the clauses are a system of inference rules. Their approach is inspired by natural deduction proof theory and defines the semantics of programs as inductively defined sets. They model non-ground answer substitutions directly, not as the set of ground instances.

Their language of Generalized Horn Clauses resembles earlier proposals for permitting nested implications in clause bodies [20, 33], but they obtain a much simpler treatment of free variables by distinguishing assumptions from program clauses. Nested implication falls outside the framework of *monotone* inductive definitions (as remarked above) but programs can be understood as *partial* inductive definitions [24].

Their approach includes a new idea, similar to elimination rules in natural deduction. A predicate $p$ is inductively defined by its set of introduction rules, namely the clauses with head $p$. If we then are told that $p$ happens to be true, then the body of some introduction rule must also be true. This gives a form of Negation as Failure and non-monotonic reasoning. Aronsson et al. [4] describe the language in more detail and discuss a prototype implementation.

### 4.3 The language Loglisp

Loglisp, by Robinson and Sibert, is one of the earlist attempts to combine logic and functional programming [39, page 400]:

> Our own early attempts (as devoted users of Lisp) to use Prolog convinced us that it would be worth the effort to create *within* Lisp a faithful implementation of Kowalski's logic programming idea. . . . We set out to honor the principle of the separation of logic from control (no CUT, no preferred ordering of assertions within procedures nor of atomic sentences within hypotheses of assertions) by making the logic programming engine 'purely denotative'.

Loglisp is not completely pure. The Logic component can invoke arbitrary Lisp functions. A more fundamental problem is the treatment of uninstantiated variables in function calls. Loglisp leaves such variables unchanged during expression reduction, so the result can depend on the order in which goals are solved. Dincbas and van Hentenryck show how this leads to anomalies [15].

But Loglisp's Horn clause interpreter — thanks to a form of *best-first search* — is complete. Although Prolog's depth-first search strategy is incomplete, this only matters if the search space is infinite, when we must be prepared to give up after a finite time. So a call to the Logic interpreter specifies how many solutions to find before stopping. We could say that Loglisp views the clauses as an inductive definition of solution sets; the Lisp half operates on lists of solutions from the Logic half. This resembles Prolog's `setof` predicate.

Robinson's later work [38] aims to integrate the functional and logic components using a single reduction semantics for both.

### 4.4 A logic language with functions

Though many combined languages have appeared since Loglisp, few tackle the problem of uninstantiated variables. Equational unification, although not completely understood, seems to be the solution. Equational unification treats functions in a natural way, retaining Prolog's bidirectionality: functions can be inverted. We have implemented such a language; a similar one is Ideal [9].

**An inductive view of functions in clauses** Viewing logic programs as inductive definitions gives a framework for a logic language with functions. Recall that the inference rules in an inductive definition contain no variables. A rule containing variables is merely an abbreviation for the set of its ground instances. We extend this means of abbreviation by permitting clauses to invoke functions defined in a functional language. Such a rule abbreviates the set of its ground instances where all functions have been evaluated.

Terms may contain *constructors* and defined *functions*. Constructors, such as constants, the pairing operator, and list Cons, generate what amounts to a Herbrand universe. Functions denote operations over this universe. Computable values are elements of this universe. Solution terms need not be ground, but should contain no defined functions. (Henceforth we shall just say 'function', not 'defined function'.)

For example, suppose $f$ and $g$ are functions. The clause

$$p(f(X), Y) \leftarrow p(X, g(Y))$$

stands for the set of instances

$$p(u, y) \leftarrow p(x, v)$$

where $u$, $v$, $x$, and $y$ are values such that $u$ is the value of $f(x)$ and $v$ is the value of $g(y)$, provided the function calls terminate.

In the first-order logic view of logic programming, programs in this kind of language are viewed as theories in Horn Clause logic plus equality. Severe restrictions are imposed on equality so that programs can be executed. The equalities must form a term rewriting system with strong properties: they must be confluent, left-linear, and terminating. In short, the equalities must take the form of function definitions. The model theory of Horn Clauses with equality, as developed by Goguen and Meseguer [22], conveys no clear picture of what their programs compute.

An inductive definition has an intuitive reading as a process generating a set of results using some rules. The function definitions help to generate the set of ground rules. An alternative picture: function evaluation is interleaved with rule application. An implementation must 'guess' suitable instances using some sort of unification. Resolving the goal $p(a, g(b))$ against $p(f(X), Y)$, requires solving the equations $a = f(X)$ and $g(b) = Y$ by unification in the equational theory of the functions $f$ and $g$.

The unification process can be seen as symbolic evaluation of the defined functions. Suppose that the list append is defined as follows:

$$\mathrm{app}([], V) \to V$$
$$\mathrm{app}([X \mid U], V) \to [X \mid \mathrm{app}(U, V)]$$

The goal $\mathrm{app}(U, V) = [a, b]$ calls the function app with uninstantiated arguments. Solving the goal requires unifying $[a, b]$ with $\mathrm{app}(U, V)$. The list $[a, b]$ is already a construction; but $\mathrm{app}(U, V)$ must be rewritten, and this instantiates $U$.

- $U = []$, the first rewrite reduces $\mathrm{app}(U, V)$ to $V$, giving the solution $V = [a, b]$.
- $U = [X_1 \mid U_1]$, for new variables $X_1$ and $U_1$, reduces $\mathrm{app}(U, V)$ to $[X_1 \mid \mathrm{app}(U_1, V)]$. Unification sets $X_1 = a$ and leaves the problem of unifying $[b]$ with $\mathrm{app}(U_1, V)$. There are two subcases.

- $U_1 = []$ gives the solution $U = [a]$ and $V = [b]$.
- $U_1 = [X_2 \mid U_2]$ sets $X_2 = b$, where we must unify $[]$ with $\mathrm{app}(U_2, V)$.
  - $U_2 = []$ gives the solution $U = [a, b]$ and $V = []$.
  - $U_2 = [X_3 \mid U_3]$ terminates the search because $[]$ and $[X_3 \mid \mathrm{app}(U_3, V)]$ cannot be unified.

This is essentially *narrowing*, a special case of the paramodulation rule used in unification algorithms for suitable equational theories [46]. There are several implementations [19, 27] and many variations. We have chosen a form of *lazy narrowing* — where a function's arguments are evaluated only when necessary — in the hope of postponing the discovery that a variable is uninstantiated. Fribourg has dealt with the strict case [18].

**Treatments of negation and 'cut'** The *cut*, written (!), curtails backtracking. Cuts can speed the search exponentially by pruning redundant parts of the search space. Cuts also compensate for implementation deficiencies, preventing the 'trail' of choice points from using up too much store. Cuts are by far the commonest impurity, and are often used needlessly. How can we do without them?

Negation as Failure can be defined through cut [13]:

```
not(P) :- call(P), !, fail.
not(P).
```

Conversely, Clocksin and Mellish recommend replacing cuts by negations whenever possible. So a pure language must include a clean treatment of Negation as Failure. This line of research is orthogonal to our own; see Minker [34].

Cuts are often used when expressing functional dependence, forcing the nondeterministic PROLOG machine to behave deterministically. By having functions in our language, we reduce the need for cuts; indeed, our prototype interpreter dynamically inserts cuts when evaluating functions.

Predicates should not be confused with boolean-valued functions. Although predicates can be represented by their characteristic functions, few logical systems formally identify them.[5] The booleans `true` and `false` are symmetric while success and failure are not. Whether a boolean function returns `true` or `false`, it has terminated successfully. But failure conveys less information than success, and may not happen in finite time.

Simple tests, such as arithmetic comparisons, should be boolean functions rather than predicates. Testing whether a boolean expression equals `false` gives

---

[5] The main formal system that does is classical higher-order logic. Intuitionistic higher-order logic identifies predicates with propositional functions, but the corresponding set of truth values is not a Boolean algebra.

a kind of negation. Our language does not allow conditional equations. The conditional rewrite rule

$$a = b \text{ if } p$$

where the condition $p$ is a predicate, says nothing about $a$ when $p$ does not hold. Instead we prefer a conditional expression controlled by a boolean expression $c$:

$$a = \text{if } c \text{ then } b_1 \text{ else } b_2$$

## 5 A Prototype of the language

This implementation was developed (by A. W. Smith) to investigate what was possible and what was desirable in such a language. It was written as an interpreter in PROLOG to carry over such features as the parser and backtracking mechanism. We have not investigated how low-level techniques for logic and functional languages might be integrated.

A program file contains clauses, function rewrite rules, type declarations, operator declarations and comments. The system is similar to a simple PROLOG interpreter but uses a lazy semantic unification algorithm. The syntax of definite clauses follows that of Edinburgh PROLOG. Built in predicates include `true`, which always succeeds; `fail`, which always fails; and $A = B$, which succeeds if $A$ and $B$ unify semantically. PROLOG's (extralogical!) output predicates `nl` and `write` are also provided.[6]

### 5.1 Rewrite Rules

Function definitions are given as a series of rewrite rules of the form

$f(expr_1, expr_2, \ldots)$ `-->` $expr$ .

For example, the append function for list is

```
app([], V)      ->> V.
app([X|U], V)   ->> [X | app(U,V)].
```

Answers to queries are returned as in PROLOG. When a solution is found, the system returns bindings for the variables. Further answers may be elicited using the '**;**' key. Example:

---

6 Because of iterative deepening, the search may pass through the same point several times and will repeat the output on each occasion.

```
| ?- solve(app(U,V)=[1,2]).
    U=[]
    V=[1,2] ;

    U=[1]
    V=[2] ;

    U=[1,2]
    V=[] ;
no
```

To ensure that the rewrite rules form a confluent system, certain restrictions are imposed. Though these are not necessary conditions, they are sufficient and are, furthermore, not unreasonable for a functional language:

**The Constructor Discipline.** The names are divided into two disjoint classes — *functions* and *constructors* — depending on whether there are rewrite rules for that name. No function should appear in the arguments of the left side of any rule.

**Left Linearity.** No variable should appear more than once in the left side of a rule.

**Term Rewriting.** All the variables appearing in the right side of a rule should also appear in the left side of the rule.

**Non-overlapping.** No two rule left sides should be unifiable with each other. (The rules should be mutually exclusive.)

If any condition is broken, the offending rewrite rule is identified in a warning. The system does not prevent the user defining and using non-confluent rules, but it contains optimisations based on the assumption that the conditions are obeyed. The interpreter assumes the constructor discipline holds, so defined functions within the left side of a rule are treated like constructors. Left linearity enables the interpreter to omit the occurs check when it unifies the function with the left side of a rule in the application of that rule. Rules which are not left linear may cause the system to loop or crash.

Finally, the assumption that rules do not overlap allows the interpreter to reduce the search space. If a function application is rewritten without instantiating any variables, then no other rule can be applicable unless it overlaps with the first. The effect is like cut — upon backtracking, no other rules need be tried.[7]

The system also checks whether the rewrite rules exhaust all possible values for the arguments. This check is rather involved. The approach is to consider the tuples formed by the arguments of the left sides of the rules. A tuple of variables is then successively instantiated (to terms of the appropriate type) so that it does not unify with each of the rule tuples taken in turn.

---

[7] A clever user could make use of this final optimisation to provide a default case for a function. This would work only for ground arguments.

### 5.2 Type Declarations

The built-in types include `int` and `bool`. Type operators such as `list` can be declared, permitting types such as `list(int)` and `list(list(bool))`. Polymorphic types like `list(A)` are permitted, as well as function types with the following syntax:

   $[type_1, type_2, \ldots]$ `=>>` $type$

Type declarations must be given for all predicates, constructors and functions. Predicate type declarations take the form

   `pred` $p(type_1, type_2, \ldots)$ .

where $type_1, type_2, \ldots$ are the types of the arguments. The type may be polymorphic: a general list appending predicate could be declared by

   `pred append(list(A), list(A), list(A)).`

The types of constructors are declared by

   `constructors` $type$ `=>` $con_1, con_2, \ldots$.

Here $con_1, con_2, \ldots$ are made up from a constructor for the given $type$, applied to type arguments. For example, the constructors for the natural numbers (type `nat`) are `zero` and `s`, where `s` must be applied to an argument of type `nat`.

   `constructors nat => zero, s(nat).`

The constructors for the polymorphic type `list(A)` are given as follows (we can use standard PROLOG list syntax, denoting PROLOG's list constructors):

   `constructors list(A) => [], [A|list(A)].`

The type of a function is declared as follows:

   `function` $f$ $(type_1, type_2, \ldots)$ `=>>` $type$.

For example:

   `function mult(nat,nat) =>> nat.`
   `function app(list(A),list(A)) =>> list(A).`

The type checking scheme follows Mycroft and O'Keefe [35] with a straightforward extension to include function rewrite rules.

A typeless version of the language could be envisaged. Type information is required by the built in equality function.

### 5.3 The equality function 'eq'

The equality function `eq` returns `true` or `false`. It behaves something like a set of rewrite rules, testing all possible combinations of constructors for a given type. The equality test is defined for each (non-function) type.

For example, the equality test for lists behaves something like the following function:

```
    [] listeq []    ->> true.
    [] listeq [A|As] ->> false.
[A|As] listeq []    ->> false.
[A|As] listeq [B|Bs] ->> (A eq B) and (As listeq Bs).
```

Here `and` denotes Boolean conjunction, while `eq` is an equality test for the list's element type.

The test is more efficient than suggested above, for `X eq t` gives `true` while instantiating `X` to `t`. When returning `false`, the equality test instantiates its arguments to all possible pairs of different constructors.

### 5.4 Other features

**Arithmetic** To perform integer arithmetic there are built in integer valued functions `+`, `-`, `*`, `div`, `mod`, and `abs` with the obvious meanings. These use the arithmetic routines of the host language but behave as if they are defined by an infinite collection of rewrite rules such as `0+0->>0`, `0+1->>1`, `-1+0->>-1`, etc. For example, the goal `15=X+Y` succeeds infinitely often, finding all pairs of numbers that sum to 15. This may seem odd, but the implementation aims to be complete. The PROLOG goal `15 is X+Y` typically results in an error message.

The relations `>`, `<`, `>=`, and `=<` are boolean valued functions — not predicates — since they are decidable.

Integers have type `int`; booleans have type `bool`.

**Higher-order functions** The built in function `apply` returns the value obtained by applying the function contained in its first argument to the list of arguments contained in its second argument. Thus `apply(+,[1,2])` returns 3. Since there is no higher-order unification, when `apply` is rewritten, the first argument must be normalisable to a function or a $\lambda$-expression.

The user can define higher-order functions:

```
map(F, [])    ->> []
map(F, [H|T]) ->> [apply(F,[H]) | map(F,T)]
```

The value of `map(lambda([X],X * X),[1,2,3])` is `[1,4,9]`.

The justification of higher-order functions requires extending the theory of narrowing to allow function variables in rewrite rules. This appears straightforward if function variables are not allowed in goals. The full incorporation of function variables would require higher-order equational unification. Our experience with higher-order unification shows that it can be effective in simple cases [36], but it also suggests that ambitious applications are impractical.

**Descriptions** Descriptions, or $\eta$-terms, call the predicate level from the function level. The term `eta(X,G)` means 'some `X` such that `G`', and returns some value of variable `X` that makes goal `G` succeed. Remaining values are found on backtracking. IDEAL has a similar feature [9, page 90].

Descriptions cause expressions to be nondeterministic, violating the separation of concerns into functions (deterministic) and predicates (nondeterministic). They pose interesting but very difficult semantic questions.

## 6 Operation of the prototype

The interpreter is essentially PROLOG with a modified unification algorithm to allow defined functions within terms. The unification algorithm is similar to that given by Martelli, Moiso and Rossi [31]. It effectively uses a selection strategy for narrowing described as 'outer narrowing' by You [45, 46]. You describes a matching algorithm; we have extended this to a unification algorithm but have not attempted a proof of correctness. The occurs check during unification could perhaps be omitted by allowing cyclic expressions to denote fixedpoints.

### 6.1 Unification

The effect of the unification algorithm can be described as follows. In unifying two terms $A$ and $B$

1. If either term is a variable, then instantiate that variable if permitted by occurs check (see below). Unification is lazy: the other term may contain function terms which are not rewritten now (maybe never).
2. If both are constructor terms, then if they have different principal constructors the unification fails, otherwise the two sets of arguments are unified.
3. Otherwise one term must be a function application. It is rewritten by narrowing (see later) to some new term, which must unify with the other term.

Only (3) distinguishes this from ordinary unification.

**Rewriting** A function application is rewritten using a rewrite rule from the function's definition. In a functional language this is achieved by matching the term with the left side of the rule. In a logic language, however, there may well be logic variables within the term; some instance of these variables might be needed to apply the rule. A narrowing step consists of unifying a term with the left side of a rule and replacing it with the right side.

If there are nested function applications, which should be chosen for narrowing? The brute-force approach would be to try every possible occurrence. This will certainly find all possible solutions but could produce the same solution many times. A practical selection strategy would chose a single occurrence. The obvious choices are the innermost or outermost occurrence.

The innermost strategy can be shown to be complete provided the rewrite system is confluent and terminating and the functions are exhaustively defined [18]. However, an innermost strategy is eager and one of the design objectives of the language was that it should be lazy.

The outermost strategy is incomplete (see also You [46]):

$$f(W, a) \to a \qquad f(a, b) \to b$$

Given the term $f(f(X, Y), Z)$, the innermost strategy can produce both $a$ and $b$, while the outermost strategy only produces $a$. The problem arises because $f(X, Y)$ can only be unified syntactically with the variable $W$ in the first rule. However if it were rewritten first to $a$ the second rule could be used as well — in other words $a$ and $f(X, Y)$ unify semantically when applying the second rule.

The solution is to use an outermost selection strategy (allowing laziness), but when narrowing, the unification between the term and rewrite rule should be semantic rather than syntactic. This differs from the usual definition of narrowing.

An advantage of using PROLOG as the host language is that the logic variables can be used as pointers. When a variable is instantiated, that value is propagated with no need to make substitutions explicitly. However the value of the pointer cannot be reassigned when a function application is rewritten. The same term could be evaluated repeatedly, giving call-by-name. To get call-by-need, function applications are represented by a structure containing a new variable to point to the rewritten value.

The unification algorithm is lazy and so only performs rewriting when required. However, the user requires answers in normal form. Thus after the system has found a solution it normalises the answer substitutions by repeatedly rewriting any defined function terms in the substitutions. The same rewriting algorithm is used as in unification. (This normalisation is too eager; in consequence, infinite data structures cannot be displayed as results, although they may take part in computations.)

**The occurs check** When a variable is to be unified with a term, it must be checked not to occur within the term. In the presence of rewrite rules an occurrence within a defined function might disappear: thus $X$ and $0 + X$ unify semantically.

The occurs check does not fail just because there is a syntactic occurrence of the variable within the term. Instead, it copies the term, replacing by a new variable each function application containing an occurrence, adding these as new disagreement pairs to (eventually) solve. If an occurrence of the variable is encountered other than in a function application, then the check fails. The new term built in this way will not contain any occurrences of the variable.

Without this extended occurs check, cases involving occurrence would have to be solved using the full unification algorithm, which would be slower.

**Search strategy** The search for a solution uses *depth-first iterative deepening* [41]. During each iteration the search is cut off if it exceeds the given limit. At the end of the iteration, if the search has been cut off at any point, then the limit is increased and the next iteration started. When a solution is found, its depth is checked to be within the range of the current iteration — to prevent a solution being returned several times.

The depth is incremented when either rewrites or clauses are applied. Thus the search is complete at both the function and clause level.

Iterative deepening is complete, straightforward to implement, and gives a sensible compromise between time and space efficiency. Perhaps some more sophisticated strategy would be more efficient, while retaining completeness.

### 6.2 Example: An Eight Queens Program

This eight queens program begins by introducing the type `list`. Predicate `upto` generates a range of numbers, while `queens` generates boards of non-attacking queens. The membership test is a function (`mem`) since its negation is used. Predicate `safe` could also be a boolean function.

```
constructors list(A) => [],[A|list(A)].

pred queens(int,list(int)).
queens(N,[Q|B]) :- (N>0)=true, queens(N-1,B), upto(1,8,Q),
                   mem(Q,B)=false, safe(1,Q,B).
queens(0, []).

pred upto(int,int,int).
upto(M,N,M) :- (M=<N)=true.
upto(M,N,K) :- (M<N)=true, upto(M+1,N,K).
```

```
    function mem(A,list(A)) =>> bool.
    mem(X,[]) ->> false.
    mem(X,[Y|Ys]) ->> (X eq Y) or mem(X,Ys).

    pred safe(int,int,list(int)).
    safe(_,_,[]).
    safe(I,Q,[Q1|B]) :- (I eq abs(Q-Q1))=false, safe(I+1,Q,B).
```

Our interpreter is slow, taking six minutes to find the first three solutions.

```
| ?- solve(queens(8,B)).
    B=[4,2,7,3,6,8,5,1] ;

    B=[5,2,4,7,3,8,6,1] ;

    B=[3,5,2,8,6,4,7,1]
yes
```

### 6.3 Example: A Propositional calculus theorem prover

Wang's Algorithm for the propositional calculus [32] works by constructing a backwards proof using the rules of the sequent calculus [21]. The following program also constructs a proof tree. Each label names some sequent calculus rule, such as `andl` for ∧-left. The subtrees represent proofs of the premises of the rule.

This program demonstrates function inversion. The function `sizeof` computes the size of proof trees. In PROLOG this function must be coded by two different predicates, depending on whether it is used to compute the size or (backwards) to generate trees of a given size.

The program begins with PROLOG infix declarations. Types of formulae, labels, and trees are declared. Note that a tree node contains a label and a list of subtrees. The remainder of the program is basically PROLOG. The program appears in Figure 1.

The first sample execution demonstrates programming with non-ground data. Here we construct non-ground trees of size 5, then instantiate them to valid proof trees. This generates theorems whose proofs have size 5. A similar program in PROLOG crashed due to a cycle (no occurs check).

The output has been beautified by indenting and by shortening internal variable names.

**Fig. 1.** A Program for Wang's Algorithm

```
/*logical connectives are constructors of type form. */
op(5,fy,~).   op(10,xfy,&).   op(20,xfy,\/).   op(30,xfy,-->).
constructors form => &(form,form), \/(form,form), ~(form),
                       -->(form,form), p,q,r.

/*labels of proof trees*/
constructors label => basic(form),andl,andr,orl,orr,notl,notr,
        impl,impr,iffl,iffr.
constructors tree => node(label, list(tree)).

function sizeof(tree) =>> int.
function sizeoflist(list(tree)) =>> int.
sizeof(node(L,Ts)) ->> sizeoflist(Ts) + 1.
sizeoflist([]) ->> 0.
sizeoflist([T|Ts]) ->> sizeof(T) + sizeoflist(Ts).

/*delmem(X,Ys,Zs) finds and removes X from Ys giving Zs */
pred delmem(A,list(A),list(A)).
delmem(X,[X|Xs],Xs).
delmem(X,[Y|Ys],[Y|Zs]) :- delmem(X,Ys,Zs).

/*common(As,Bs,B) when B is a common element of As and Bs*/
pred common(list(A),list(A),A).
common(As,[B|_],B) :- delmem(B,As,_).
common(As,[_|Bs],B) :- common(As,Bs,B).

/*proof(left formulae,right formulae,proof tree)  */
pred proof(list(form), list(form), tree).

proof(As,Bs,node(basic(B),[])) :- common(As,Bs,B). /*0 subproofs*/

/*1 subproof*/
proof(As,Bs,node(notr,[T])) :- delmem(~B,Bs,Ds),proof([B|As],Ds,T).
proof(As,Bs,node(andl,[T])) :- delmem(A1&A2,As,Cs), proof([A1,A2|Cs],Bs,T).
proof(As,Bs,node(orr,[T]))  :- delmem(B1\/B2,Bs,Ds),proof(As,[B1,B2|Ds],T).
proof(As,Bs,node(notl,[T])) :- delmem(~A,As,Cs), proof(Cs,[A|Bs],T).
proof(As,Bs,node(impr,[T])) :- delmem(B1-->B2,Bs,Ds),
        proof([B1|As],[B2|Ds],T).

/*2 subproofs*/
proof(As,Bs,node(andr,[T1,T2])) :- delmem(B1&B2,Bs,Ds),
        proof(As,[B1|Ds],T1), proof(As,[B2|Ds],T2).
proof(As,Bs,node(orl,[T1,T2])) :- delmem(A1\/A2,As,Cs),
        proof([A1|Cs],Bs,T1), proof([A2|Cs],Bs,T2).
proof(As,Bs,node(impl,[T1,T2])) :- delmem(A1-->A2,As,Cs),
        proof(Cs,[A1|Bs],T1), proof([A2|Cs],Bs,T2).
```

```
?- solve((5=sizeof(T), proof([],[B],T))).
    T=node(andr,[node(impr,[node(basic(_1),[])]),
                 node(impr,[node(basic(_2),[])])])
    B=(_1-->_1)&(_2-->_2) ;

    T=node(impr,[node(andr,[node(basic(_1),[]),
                            node(impr,[node(basic(_2),[])])])])
    B=_1-->_1&(_2-->_2) ;

    T=node(impr,[node(andr,[node(basic(_1),[]),
                            node(impr,[node(basic(_1),[])])])])
    B=_1-->_1&(_2-->_1) ;

    T=node(impr,[node(andr,[node(basic(_1&_2),[]),
                            node(andl,[node(basic(_1),[])])])])
    B=_1&_2-->(_1&_2)&_1
yes
```

The program can also prove theorems and report the size of the proof.

```
?- solve((proof([],[p & (q & r)  -->  (p & q) & r],T), N=sizeof(T))).
    T=node(impr,[node(andl,[node(andl,
        [node(andr,[node(andr,[node(basic(p),[]),
                               node(basic(q),[])]),
                    node(basic(r),[])])])])])
    N=8
yes
```

## 7 Conclusions

We have criticised logic programming as a declarative paradigm. What can be done to make pure logic programming practical?

To justify the Closed World Assumption, we propose that logic programs should be viewed as inductive definitions, not as first-order theories. Some people refuse to abandon the dream of programming in first-order logic. But we have to ask whether this dream is possible — even whether it is desirable. The first-order paradigm does not deal adequately with negation in databases, and seems to be an unreliable guide in research on program correctness and language design. Inductive definitions are more fundamental than first-order logic, and perhaps easier to understand.

Uses of PROLOG's cut can be largely eliminated by providing functions and negation. Two forms of negation can be mathematically justified: testing whether a boolean expression returns `false`, and restricted forms of Negation as Failure. Concepts such as 'stratified program' are best understood from the perspective of inductive definition. Future languages should provide means for organizing a program into a suite of inductive definitions, since stratification will not hold by accident.

The PROLOG goal var(X) tests whether or not the variable X is bound. This clearly has no logical meaning. Some uses of var can be eliminated by allowing functions to accept non-ground arguments. But there are some difficult examples, such as writing a polymorphic type checker [11]. This is natural to write in PROLOG, since it uses unification and search. When the type has been inferred, certain of its variables must be labelled as 'generic', and such manipulation of logical variables must use var.

Such uses of var serve not to make up for language deficiencies, but to exploit global properties of the program. Similarly, it is hard to eliminate certain instances of cut. We must either retain such impurities in our languages, or be prepared to tolerate some inconvenience.

Input/output is the greatest challenge. Pure approaches to input/output constitute much of the research in functional programming, and perhaps could be applied to the functional part of our language. Most approaches involve continuations, so our extended logic languages must provide higher-order functions.

Our prototype is far too slow for programmers. But the authors of IDEAL, a similar language, claim outstanding efficiency [5, 8]. Their system translates functions into PROLOG clauses, and then into a modified Warren Abstract Machine. It is incomplete due to depth-first search, but presumably there could be a version using iterative deepening. An OR-parallel machine such as DelPhi [12] could support such languages in future. Functions make explicit the granularity for OR-parallelism: evaluation is deterministic while search is not.

*Acknowledgements.* This work was supported by the Alvey Diamond project: SERC grants GR/E/02369 and GR/F/10811. William Clocksin, Martin Hyland, Tobias Nipkow, Andrew Pitts, Peter Schroeder-Heister, Lincoln Wallen, David Wolfram, and the referees commented on drafts of this paper.

This article was processed using the LaTeX macro package with ICM style